\documentclass[aps,prev,12pt,preprintnumbers,floatfix,nofootinbib]{revtex4-1}
\usepackage{graphicx}
\usepackage{bm}
\usepackage{times}
\usepackage{amsmath}
\usepackage{amsfonts}
\usepackage{amssymb}
\usepackage{slashed}
\usepackage{color}
\usepackage{epstopdf}
\usepackage{subfigure}

\def\br{\begin{eqnarray}}
\def\er{\end{eqnarray}}
\def\be{\begin{equation}}
\def\ee{\end{equation}}

\def\({\left(}
\def\){\right)}

\newcommand \beq { \begin{eqnarray} }
\newcommand \eeq { \end{eqnarray} }
\newcommand \beqq { \begin{equation} }
\newcommand \eeqq { \end{equation} }

\begin{document}

\title{The Muon Anomalous Magnetic Moment in the Reduced Minimal 3-3-1 Model}
\vspace{2cm}
\author{Chris Kelso$^a$}
\author{P.R.D. Pinheiro$^b$}
\author{Farinaldo S. Queiroz$^c$}
\author{William Shepherd$^c$}
\vspace{2cm}
\affiliation{$^a$ Department of Physics and Astronomy, University of Utah, Salt Lake City, UT 84112, USA\\ 
$^b$ Grupo de F\'{i}sica Te\'{o}orica Jayme Tiomno, S\~{a}o Lu\'{i}s, Maranh\~{a}o, Brazil.\\
$^c$Department of Physics and Santa Cruz Institute for Particle Physics
University of California, Santa Cruz, CA 95064, USA}

\vspace{2cm}
\begin{abstract}
We study the muon anomalous magnetic moment $(g-2)_{\mu}$ in the context of the reduced minimal 3-3-1 model recently proposed in the literature. In particular, its spectrum contains a doubly charged scalar ($H^{\pm \pm}$) and gauge boson ($U^{\pm \pm}$), new singly charged vectors ($V^{\pm}$) and a $Z^{\prime}$ boson, each of which might give a sizeable contribution to the $(g-2)_{\mu}$. We compute the 1-loop contributions from all these new particles to the $(g-2)_{\mu}$. We conclude that the doubly charged vector boson provides the dominant contribution, and by comparing our results with the experimental constraints we derive an expected value for the scale of $SU(3)_L\otimes U(1)_N$ symmetry breaking $v_{\chi} \sim 2$~TeV. We also note that, if the discrepancy in the anomalous moment is resolved in the future without this model then the constraints will tighten to requiring $v_\chi > 2.7$ TeV with current precision, and will entirely rule out the model if the expected precision is achieved by the future experiment at Fermilab.

\vspace{2cm}    
\end{abstract}

\maketitle

\section{Introduction}

Now that the Higgs discovery has completed the Standard Model (SM), the muon anomalous magnetic moment, one of the most precisely measured quantities in particle physics, seems to be the most compelling ``discrepancy" between
theory and experiment. A long standing $2-3\sigma$ difference from the SM predicted value has been observed \cite{PDG}. This deviation has triggered
numerous speculations about its possible origin, and the increased
experimental precision over time inspired a multitude of new theoretical efforts which led to
a substantial improvement of the prediction of the muon magnetic moment, generally written in terms of $a_{\mu}=(g_{\mu}-2)/2$. The comparison between experiment and the
SM is a sensitive test of new physics. At present, both measurement and
theory quote similar uncertainties and $a_\mu$ has been used to used to constrain
standard model extensions. In fact, the difference, $a^{exp}_{\mu}-a^{SM}_{\mu}= (296 \pm 81) \times 10^{-11}$, which corresponds to a $3.6\sigma$ discrepancy, imposes quite stringent bounds on many particle physics models \cite{fermilabproposal}. A worldwide effort is underway to reduce this uncertainty, with the ultimate hope of either strengthening evidence for the presence of new physics or refuting the discrepancy through more accurate SM calculations. It is important to remind ourselves that this $3.6\sigma$ deviation is reduced to $2.4\sigma$ if one uses $\tau$ data in the hadronic contributions \cite{PDG}. 

One could attempt to interpret this difference as coming from theoretical uncertainties. The SM prediction for $a_{\mu}$ is generally divided into three parts: electromagnetic (QED) corrections, electroweak (EW) corrections, and hadronic contributions. The QED part includes all photonic and leptonic ($e,\mu,\tau$) loops, and the EW involves $W^{\pm},Z$ and Higgs graphs. Hadronic contributions are only possible through the couplings of hadronic matter to color-neutral bosons of the SM, with the dominant contributions coming from the couplings to the photon. The two main contributions are the hadronic vacuum polarization and the hadronic contribution to the light-by-light scattering graph. The hadronic corrections give rise to the main theoretical uncertainties, but it is expected that those uncertainties will be reduced in the foreseeable future due to improvement in lattice QCD and hadronic data \cite{g2muontheory1,g2muontheory2,g2muontheory3,g2muontheory4}.

A more popular approach to this anomaly is to treat it as evidence for new physics, as investigated by various authors for multitudes of models \cite{gmuonmodels0,gmuonmodels1,gmuonmodels2,
gmuonmodels3,gmuonmodels4,gmuonmodels5,gmuonmodels6}. In particular, the $a_\mu$ anomaly has also been investigated in the context of 3-3-1 models \cite{g2muon331_1,g2muon331_2,g2muon331_3,g2muon331_4}. Here we will focus on a specific realization of the 3-3-1 gauge symmetry with a smaller fermion and scalar sector known as the reduced minimal 3-3-1 model \cite{RM331,RM331_2}. In addition to having fewer matter fields, this model also features doubly-charged vector and scalar bosons. These lead to interesting new phenomenology. As an example, because of these new charged scalars and gauge bosons, this model might be able to reproduce the now faint $H\to\gamma \gamma$ excess \cite{HRM331}.  In this work we aim to explore the implications of the $a_{\mu}$ anomaly on the reduced 3-3-1 model. This issue has been briefly addressed in \cite{Yue:2013qba}, considering only contributions from the singly- and doubly-charged gauge bosons. In this work, we will derive analytical expressions for all 1-loop contributions coming 
from the reduced 3-3-1 model, and draw our conclusions based on the total contribution. 

This paper is organized as follows: in Section \ref{sec1} we briefly introduce the model. Section III discusses the $a_{\mu}$ discrepancy from the SM, and in Section IV we present the $a_\mu$ predictions of the reduced 3-3-1 model. Lastly, in Section \ref{sec4} we draw our conclusions.

\section{The Reduced Minimal 3-3-1 Model and $a_\mu$}
\label{sec1}

The Reduced Minimal 3-3-1 model (hereafter RM331) is based on the $SU(3)_C \times SU(3)_L \times U(1)_N$ gauge group, and therefore the left handed fermions must appear in $SU(3)_L$ triplets. Right-handed SM fermions can either be singlets of $SU(3)_L$ or may be charged only under the generators of $SU(3)_L$ which are broken at a fairly high scale to give back the SM $SU(2)_L$.

\subsection{Fermions} 

In the RM331, the fermions are embedded in the following multiplets\footnote{There are many other 3-3-1 models which are comprised of different scalar and fermions sectors. See Refs.\cite{331versions1,331versions2,331versions3,331versions4,331versions5,331versions6,
331versions7,331versions8,331versions9,331versions10,331versions11,331versions12,
331versions13,331versions14,331versions15,331versions16,331versions17,331versions18,
331versions19,331versions20} for some of these.} 

\begin{equation}
\begin{array}{cc}
f_L=\left(\begin{array}{c}
\nu_l \\  l\\ l^c
\end{array}\right)_L \sim (1,3,0),
\end{array}
\label{leptonscontent}
\end{equation}where $ l=e,\mu ,\tau$.
\begin{eqnarray}
\label{quarks}
&&
\begin{array}{c}
 Q_{1L}=\left(\begin{array}{cc}
 u_1 \\ d_1 \\ J_1
\end{array}\right)_L \sim (3,3,+\frac{2}{3})
\end{array}\,\,,\,\,
\begin{array}{c}
Q_{iL}=\left(\begin{array}{cc}
 d_i \\ -u_i \\ J_i^{\prime}
\end{array}\right)_L \sim (3,3^*,-\frac{1}{3}),
\end{array}
\nonumber \\
&&
\begin{array}{ccc}
u_{1R} \sim(3,1,+\frac{2}{3}); & d_{1R}
\sim(3,1,-\frac{1}{3}); & J_{1R}
\sim(3,1,+\frac{5}{3}), \label{quarkcontent}
\end{array}
\nonumber \\
&&
\begin{array}{ccc}
u_{iR} \sim(3,1,+\frac{2}{3}); & d_{iR}
\sim(3,1,-\frac{1}{3}); & J^{\prime}_{iR}
\sim(3,1,-\frac{4}{3}),
\end{array} 
\end{eqnarray}
where $i=2,3$. Here $J$ fields are new heavy quarks. Note that the right-handed SM leptons are actually charged under $SU(3)_L$, but the right-handed SM quarks are not.

In parentheses we have shown the quantum number of these field under the gauge group $SU(3)_C \times SU(3)_L \times U(1)_N$. Due to chiral anomaly cancellation conditions, the quark families must be placed in different representations of $SU(3)_L$, as shown in Eq.(\ref{quarks}). The anomaly cancellation conditions also actually require the existence of a minimum of 3 fermion families in this mode. It should be noted that the electric charges of the $J_{1}$ and $J_{i}$ quarks are $+5/3$ and $-4/3$, respectively, making them exotic quarks. The phenomenology of these exotics has been explored in \cite{Exoticquarks}.
\subsection{Scalars}

The scalar sector is comprised of two Higgs triplets:
\begin{equation}
\begin{array}{cccc}
\rho =\left(\begin{array}{c}
\rho^+ \\ \rho^0 \\ \rho^{++}
\end{array}\right) \sim (1,3,1);&
\chi =\left(\begin{array}{c}
\chi^- \\ \chi^{--} \\ \chi^0\end{array}\right) \sim (1,3,-1),
\end{array}
\label{scalarcontent}
\end{equation} and its interactions are described by the potential
\begin{eqnarray}
V(\chi,\rho)&=&\mu^2_1\rho^\dagger\rho+
\mu^2_2\chi^\dagger\chi+\lambda_1(\rho^\dagger\rho)^2+\lambda_2(\chi^\dagger\chi)^2
\nonumber \\ & &\mbox{}
+\lambda_3(\rho^\dagger\rho)(\chi^\dagger\chi)+\lambda_4(\rho^\dagger\chi)(\chi^\dagger\rho).
\label{potential}
\end{eqnarray}
This potential gives rise to the spontaneous symmetry breaking mechanism when $\rho^0$ and $\chi^0$ develop VEVs as follows,
\begin{equation}
\rho^0 \,,\, \chi^0 \rightarrow \frac{1}{\sqrt{2}}(v_{\rho\,,\,\chi}+ R_{\rho\,,\,\chi} +iI_{\rho\,,\,\chi}).
\label{VEVs}
\end{equation} The constraints on the couplings induced by our definition of the VEVs above are
\begin{eqnarray}
&&
 \mu^2_1+\lambda_1 v^2_\rho+\frac{\lambda_3 v^2_\chi}{2}=0,\nonumber \\
 &&
 \mu^2_2+\lambda_2 v^2_\chi+\frac{\lambda_3 v^2_\rho}{2}=0.
\end{eqnarray}

This scalar sector is sufficient to induce the correct pattern of symmetry breaking, where $SU(3)_C \times SU(3)_L \times U(1)_N$ breaks into the SM gauge group. This breaking occurs when the $\chi$ scalar develops a vacuum expectation value. The second spontaneous symmetry breaking happens when $\rho^0$ component acquires a vev, $v_\rho$, breaking the SM gauge group down to $SU(3)_C \times U(1)_{Q}$. Naturally, $v_\rho$ will be identified with the SM Higgs vev. At the end of the day, we find the mass matrix in the basis $(\chi^{++}\,,\,\rho^{++})$ to be 
\begin{equation}
m^2_{++}=\frac{\lambda_4 v^2_\chi}{2}
\begin{pmatrix}
t^2\ & t  \\
t& 1 
\end{pmatrix},
\label{++massmatrix}
\end{equation}where $t=\frac{v_\rho}{v_\chi}$. Moving to the mass basis gives
\begin{equation}
m^2_{\tilde h^{++}}=0\,\,\,\,\,\,\,\mbox{and}\,\,\,\,\,\,m^2_{h^{++}}=\frac{\lambda_4}{2}(v^2_\chi + v^2_\rho),
\label{masscharged}
\end{equation}
where the corresponding eigenstates are,
\begin{eqnarray}
&&\left( 
\begin{tabular}{c}
$\tilde h^{++}$ \\ 
$h^{++} $
\end{tabular}
\ \right) = \left( 
\begin{tabular}{cc}
$c_ \alpha$ & -$s_ \alpha$\\ 
$s_\alpha$ & +$c_\alpha$
\end{tabular}
\ \right) \left( 
\begin{tabular}{l}
$\chi^{++} $ \\ 
$\rho^{++} $
\end{tabular}
\right),
\label{chargedeigenvectors}
\end{eqnarray}
with 
\begin{equation}
c_\alpha = \frac{v_\chi}{\sqrt{v^2_\chi + v^2_\rho}}\,,\,s_\alpha=\frac{v_\rho}{\sqrt{v^2_\chi + v^2_\rho}}.
\label{chargedeiggenvectors}
\end{equation}
Therefore, $\tilde h^{++}$ is a (would-be) Goldstone boson, and in the limit $v_{\chi} \gg v_{\rho}$, $h^{++} \sim  \rho^{++}$.
As for the neutral scalars $(R_\chi\,,\,R_\rho)$ we find the mass matrix,
\begin{equation}
m^2_0=\frac{v^2_\chi}{2}
\begin{pmatrix}
2\lambda_2\ & \lambda_3 t  \\
\lambda_ 3 t& 2\lambda_1 t^2 
\end{pmatrix},
\label{neutralmassmatrix}
\end{equation} which gives
\begin{equation}
m^2_{h^1}=(\lambda_1 -\frac{\lambda^2_3}{4\lambda_2})v^2_\rho\,\,\,,\,\,\,m^2_{h_2}=\lambda_2v^2_\chi +\frac{\lambda^2_3}{4\lambda_2}v^2_\rho,
\label{neutralmass}
\end{equation}with
\begin{eqnarray}
h_1=c_\beta R_\rho-s_\beta R_\chi \,\,\,,\,\,\,
h_2=c_\beta R_\chi +s_\beta R_\rho \,,
\label{neutrales}
\end{eqnarray}where $c_\beta=\cos(\beta) \approx 1-\frac{\lambda^2_3}{8 \lambda^2_2}\frac{v^2_\rho}{v^2_\chi}$ and $s_\beta=\sin(\beta) \approx \frac{\lambda_3}{2\lambda_2}\frac{v_\rho}{v_\chi}$. $h_1$ is identified as the SM higgs when $sin(\beta) \rightarrow 0$.

Counting degrees of freedom one can conclude that there should remain a doubly charged and two neutral scalars in the spectrum after symmetry breaking. The other scalars are ``eaten'' as follows: $\chi^{\pm}$ is absorbed by the gauge boson $V^{\pm}$, $\rho^{\pm}$ by $W^{\pm}$, one combination of the doubly charged scalars $\rho^{\pm \pm}$ and $\chi^{\pm \pm}$ gives rise to the massive scalar $H^{\pm \pm}$ while the other is absorbed by the doubly charged boson $U^{\pm \pm}$. Moreover, the pseudo-scalars $I_{\rho}$ and $I_{\chi}$ are eaten by the Z and $Z^{\prime}$ bosons as aforementioned. 

\subsection{Gauge Bosons}

The gauge boson masses due to this symmetry breaking are

\begin{eqnarray}
&& 
M_{W^{\pm }}^{2} =\frac{g^{2}v_{\rho }^{2}}{4} ,\,\,\,\ m^2_Z=\frac{g^2}{4c^2_W}v^2_\rho,\\
&&
M_{V^{\pm }}^{2} =\frac{g^{2}v_{\chi }^{2}}{4} , \\
&&
M_{U^{\pm \pm }}^{2} =\frac{g^{2}\left( v_{\rho }^{2}+v_{\chi }^{2}\right) }{4},\\
&&
m^2_{Z^{\prime}}=\frac{g^2 c^2_W}{3(1-4 s^2_W)}v^2_\chi.\
\label{masseschargedbosons}
\end{eqnarray} where $c_W=\cos\theta_W$, $s_W=\sin\theta_W=t/\sqrt{1+4t^2}$, $t=g_N/g$, $t_W=\tan \theta_W$, $h_W = 1-4 s^2_W$, and $\theta_W$
is the Weinberg angle. Due to the spontaneous symmetry breaking pattern we can relate the $U(1)_N$ and $U(1)_Y$ gauge couplings. Using the fact that $g/g'=c_W/s_W$ we find,
\begin{equation}
\label{landau}
\frac{g_N^2}{g^{2} }=\frac{s_W^2}{1-4s_W^2}.
\end{equation}

Therefore we have a Landau pole when $s_W^2=1/4$. Indeed, the problem of the Landau pole
indicates that the coupling constant $g_N$ diverges at sufficiently high energy scale. Since $g$ is the coupling constant of the $SU(2)_L$ group, which is embedded in the $SU(3)_L$ group, it is measured to be finite and thus cannot be driven small to satisfy Eq. \ref{landau}. In this context the Landau pole $\Lambda$ stands for the energy cutoff of the 3-3-1 symmetry \cite{341}. In the RM331, the Landau pole was found to be $\Lambda\sim 5$~TeV \cite{pole}. Therefore this model is necessarily within reach of the 14 TeV LHC. 

Since we will focus on the $a_{\mu}$ anomaly, the quark sector is largely irrelevant, contributing only at higher-loop order. Therefore we will restrict our discussion to the couplings of scalars and gauge bosons to the leptons, which contribute at one-loop order. We consider three classes of interactions.

\subsection{Charged Current Interactions}

The charged and doubly-charged current interactions predicted by the RM331 model are

\beq
{\cal L}^{CC}_{l}= \frac{g}{\sqrt{2}}\bar \nu_{a_L} \gamma^\mu 
V^l_{PMNS}e_{a_L}W^+_\mu + \frac{g}{\sqrt{2}}\overline{e^c}_{a_L}O^V\gamma^\mu 
\nu_{a_L} V^+_\mu + \frac{g}{\sqrt{2}} \overline{e^c}_{a_L} \gamma^\mu 
e_{a_L}U^{++}_\mu +h.c,\nonumber \\
\label{4.6}
\eeq

where $a=1,2,3$ with  $V^l_{PMNS}=V^{\nu \dagger}_L $ being the PMNS mixing matrix and $O^V=V^{\nu }_L$ is the matrix which diagonalizes the neutrino mass matrix. We will neglect the lepton mixings in this work because their contribution to the anomalous magnetic moment of the muon is very small. 

\subsection{Neutral Current Interactions}

The neutral currents for leptons are,
\beq
{\cal L}^{NC}_{l}&=&-\frac{g}{2c_W}\bar \nu_{a_L}\gamma^\mu \nu_{a_L}Z_\mu -\frac{g}{2c_W} 
\sqrt{\frac{h_W}{3}}\bar \nu_{a_L}\gamma^\mu \nu_{a_L}Z^{\prime}_\mu 
\nonumber \\
&&-\frac{g}{2c_W}\bar e_a\gamma^\mu \left(a_1 - b_1 \gamma^5 \right)e_aZ_\mu 
-\frac{g}{2c_W}\bar e_a\gamma^\mu \left(a_2 - b_2 \gamma^5 \right)e_aZ^{\prime}_\mu,
\label{4.8}
\eeq
where,
\beq
&&a_1=-\frac{1}{2} h_W \,\,\, , \,\,\,b_1 = 
-\frac{1}{2}\nonumber \\
&&a_2=\frac{1}{2}\sqrt{3h_W} \,\,\, , \,\,\,b_2 = 
-\frac{1}{2}\sqrt{3h_W}.
\label{4.9}
\eeq

As we will see further the $Z^{\prime}$ interactions give rise to somewhat negative sizeable contributions to the $(g-2)_\mu$.

\subsection{Scalar Interactions}

Here we present only the relevant interactions among charged leptons and scalars. They arise from effective dimension five operators, which we assume to be suppressed by the Landau pole of the theory,
\begin{equation}
\frac{G_{ab}}{\Lambda}\left(\overline{f^c_{aL}}\rho^*\right)\left(\chi^\dagger f_{bL} \right) + h.c.
\label{chargedleptonmasses}
\end{equation} producing leptons masses given by,

\begin{equation}
m_{l_a} = G_{ab}v_{\rho}v_{\chi}/(2\Lambda).
\end{equation}

Notice that, even though the leptons masses come from a non-renormalizable operator, the leptons masses end up being the same, and since $v_{\chi} \sim \Lambda$, they suffer from the same SM fine-tuning in the Yukawa couplings. From Eq.~(\ref{chargedleptonmasses}), we obtain the following interactions,

\begin{eqnarray}
{\cal L}_{l}&=&\frac{m_l}{v_{\rho}}\left(\cos(\beta) - \frac{v_\rho}{v_{\chi}}\sin(\beta) \right)\bar{l}lh_1+\frac{m_l}{v_{\rho}}\left(\sin(\beta) + \frac{v_\rho}{v_{\chi}}\cos(\beta) \right)\bar{l}lh_2 \nonumber \\ & &\mbox{}
+\sqrt{2}\frac{ m_l}{v_{\rho}}h^{--}\overline{l}P_L(l^c) + h.c
\label{leptonsscalarinteractions}
\end{eqnarray}where $l=e, \mu , \tau$ and $P_L$ is equal to $(1-\gamma_5)/2$.

\section{The Muon Anomalous Magnetic Moment}

The muon magnetic moment is related to its intrinsic spin by the gyromagnetic ratio $g_{\mu}$:

\begin{equation}
\overrightarrow{\mu}_{\mu} = g_{\mu} \left( \frac{q}{2m} \right) \overrightarrow{S}
\end{equation}

where $g_{\mu}$, within the framework of the Dirac equation, is expected to be equal to two for a structureless spin 1/2 particle. However, quantum loop corrections associated with QED, electroweak, and QCD processes lead to a deviation from this value which are parametrized in terms of $a_{\mu}= (g_{\mu}-2)/2$. The SM prediction for the $a_{\mu}$ is generally divided into three parts: electromagnetic (QED), electroweak (EW) and hadronic contributions. The QED part includes all photonic and leptonic ($e,\mu,\tau$) contributions and has been computed up to 4-loops and estimated at the 5-loop level. The EW involves $W^{\pm},Z$ and Higgs bosons, and has been computed up to three loops. The hadronic contributions are the most uncertain. The hadronic vacuum polarization is calculated and inferred either from $e^+e^- \rightarrow$ hadrons or $\tau \rightarrow$ hadrons data \cite{PDG}. The next largest uncertainty is associated with hadronic light-by-light scattering, which cannot, at present, be determined from 
data, but rather must be calculated using hadronic models that correctly reproduce the properties of QCD \cite{lightlight}. Ultimately, the final value is found to be \cite{fermilabproposal},

\begin{equation}
a_{\mu}^{SM} = (116591785 \pm 51) \times 10^{-11}.
\end{equation}

Recently, the E821 experiment has measured \cite{E821_1,E821_2,E821_3},

\begin{equation}
a_{\mu}^{E821}= (116592080 \pm 63) \times 10^{-11}.
\end{equation}
Hence, 

\begin{equation}
\Delta a_{\mu} (E821 -SM) = (295 \pm 81) \times 10^{-11},
\label{deltaa}
\end{equation}which points to a $3.6\sigma$ excess. The present theoretical error of $\pm 51 \times 10^{-11}$ is dominated by the $\pm 39 \times 10^{-11}$ uncertainty on lowest-order hadronic contribution and the $\pm 26 \times 10^{-11}$ uncertainty on the hadronic light-by-light contribution \cite{fermilabproposal}. It has been suggested that uncertainty on the lowest-order hadronic contribution could be reduced to $25 \times 10^{-11}$ with existing data and further work on the hadronic light-by-light corrections could reduce the total SM error to as little as $\pm 30 \times 10^{-11}$ \cite{fermilabproposal,g2muontheory2}.
With the proposed experimental error of $\pm 16 \times 10^{-11}$ for the experiment with improved statistics at Fermilab, the combined uncertainty for the difference between theory and experiment  might reach $\pm 34 \times 10^{-11}$, better by a factor $\sim 2.4$ than the current error \cite{fermilabproposal}. We will utilize this latter value as an approximation of the future sensitivity of this observable for our further calculations.

\begin{figure}[!h]
\subfigure[\label{scalar1}]{\includegraphics[scale=0.3]{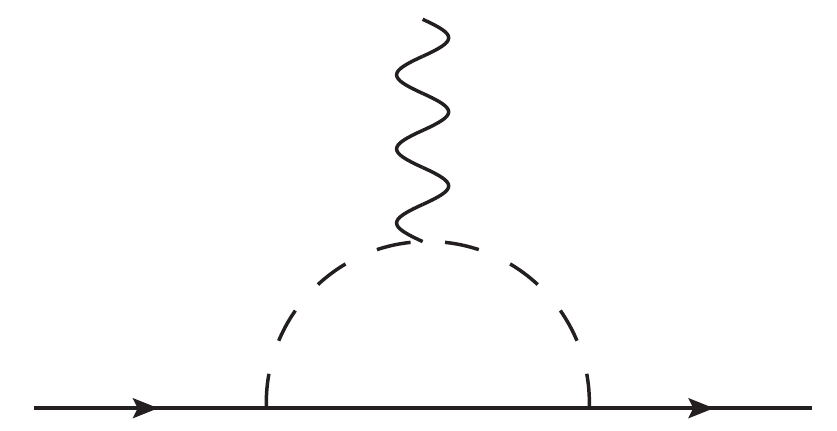}}
\subfigure[\label{scalar2}]{\includegraphics[scale=0.3]{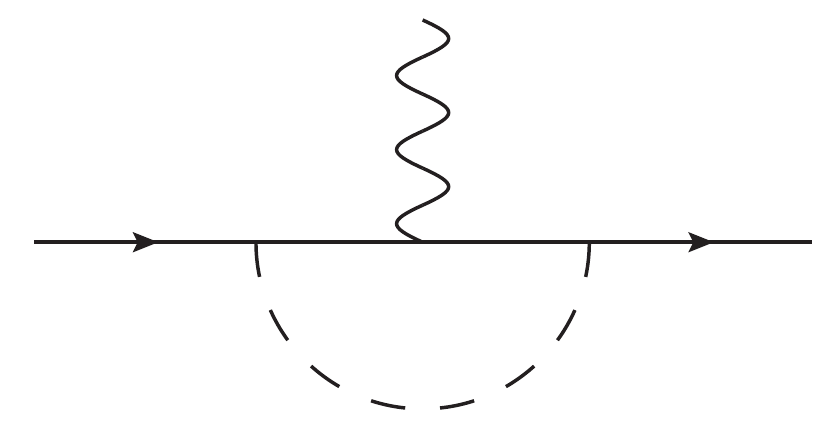}}
\subfigure[\label{vector1}]{\includegraphics[scale=0.3]{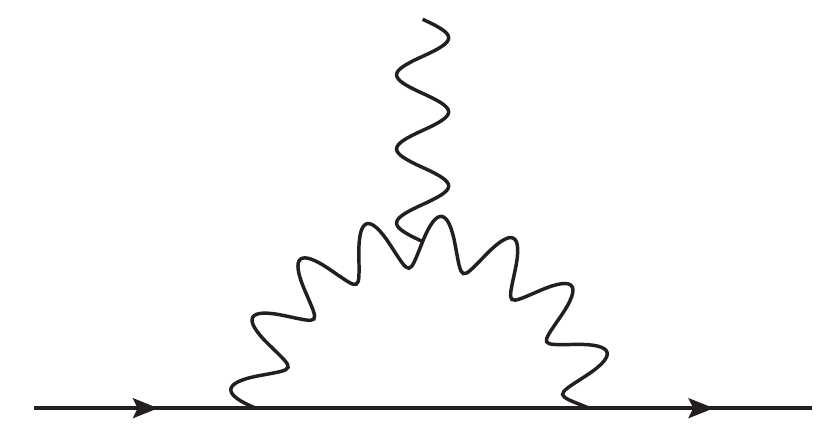}}
\subfigure[\label{vector2}]{\includegraphics[scale=0.3]{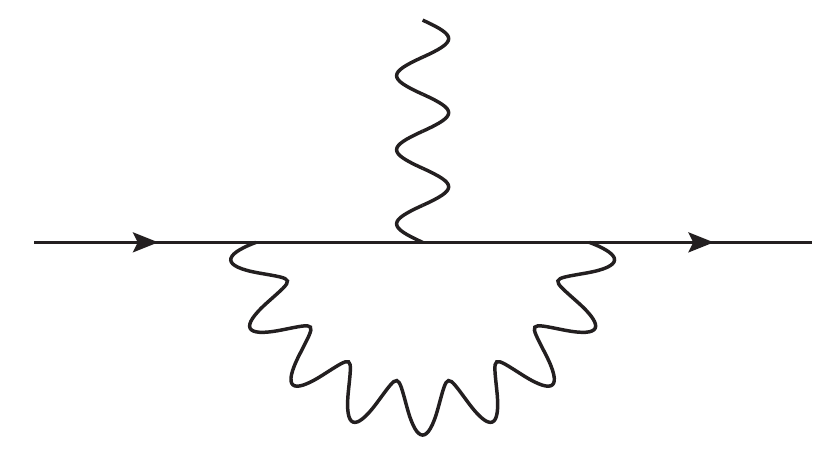}}
\caption{Feynman graphs of one-loop contributions to $a_{\mu}$. }
\label{diagrams}
\end{figure}

\section{Contributions to the $(g-2)_{\mu}$}

We now turn to consider the implications of the anomaly in $a_\mu$ from the perspective of the RM331 model. The only new particles which contribute to the muon anomalous magnetic moment at first order are the bosons, as there are no new leptons in the fermion sector of the RM331 and the quarks will only contribute at higher order. We will explore each new boson's contributions independently, and reach our ultimate conclusions based on the sum of all the contributions.

\subsection{Singly Charged Vector}

The charged vector bosons $V^\pm$ contribute to the muon anomalous moment through the diagram shown in Figure \ref{vector1}, which leads to the expressions given in Ref.~\cite{muonreport},

\begin{eqnarray}
\Delta a_{\mu} (V^{\pm}) =\frac{f^2_V}{8\pi^2}\frac{m_\mu^2}{M_V^2}\int_0^1 dx \frac{P_V(x) +P_A(x) }{\epsilon^2 \lambda^2(1-x)(1-\epsilon^{-2}x) + x }
\label{Vcontri1}
\end{eqnarray}where $f_V=g/\sqrt{2}$ is the coupling strength between the muon and the new boson given in Eq.(\ref{4.6}), with $\epsilon = m_{\nu}/m_{\mu}$,\, $\lambda = m_{\mu}/M_V$ and

\begin{eqnarray}
P_V(x) =2 x^2(1+x-2\epsilon)+\lambda^2(1-\epsilon)^2\cdot x (1-x)(x+\epsilon) \nonumber\\
P_A(x) =2 x^2(1+x+2\epsilon)+\lambda^2(1+\epsilon)^2\cdot x (1-x)(x-\epsilon).
\end{eqnarray}
The reason we have two terms in Eq.(\ref{Vcontri1}) is due to the presence of vector (V) and axial-vector (A) couplings in the muon-charged boson interaction Lagrangian in Eq.(\ref{4.6}). In the limit the mass of the singly charged boson running in the loop is much larger than the neutrino and muon masses we find,

\begin{eqnarray}
\Delta a_{\mu}(V^{\pm}) = \frac{g^2 m_{\mu}^2}{4 \pi^2 M_V^2}\left( \frac{10}{6}\right).
\label{NPGa}
\end{eqnarray}
This is the contribution of the singly charged gauge boson to the muon anomaly magnetic moment.

\subsection{Doubly Charged Scalar}

As for the doubly-charged scalar, the diagrams of Figure \ref{scalar1}-\ref{scalar2} both contribute, and we find 

\begin{eqnarray}
\Delta a_{\mu} (H^{\pm \pm}) & = & (4)\times \frac{-q_H}{4 \pi^2} \left( \frac{f_{H} m_\mu}{M_{H^{\pm \pm}}}\right)^2 \int^1_0 dx \frac{x^3-x^2}{ (\lambda x)^2 + (1-2\lambda^2)x + \lambda^2 }+\nonumber\\
&  & (4)\times \frac{-q_f}{4 \pi^2} \left( \frac{f_{H} m_\mu}{M_{H^{\pm \pm}}}\right)^2 \int^1_0 dx \frac{x^2-x^3}{ (\lambda x)^2 + (1-x)}
\label{Hppcontri}
\end{eqnarray}where $f_H=k\ v_{\chi}/(\sqrt{2}\sqrt{v_{\chi}^2+v_{\rho}^2})$ is the coupling strength between the muon and the new boson given in Eq.(\ref{4.6}), with $\epsilon = m_{\nu}/m_{\mu}$,\, $\lambda = m_{\mu}/M_{H^{\pm \pm}}$, $q_H=-2$ is the electric charge of the doubly charged scalar running in the loop, and $q_f=1$ is the electric charge of the muon in the loop. The factor of four in Eq.(\ref{Hppcontri}) is a symmetry factor due to the presence of two identical fields in the interaction term, as discussed in Ref.~\cite{MuonC}. The reason we have two integrals in Eq.(\ref{Hppcontri}) is due to the presence of two distinct diagrams. This expression simplifies to give

\begin{equation}
\Delta a_{\mu}(H^{\pm \pm})= \frac{-2}{3} \left(\frac{f_H m_{\mu} }{\pi M_{H^{\pm \pm}} } \right)^2
\label{doublyScalar}
\end{equation}

Note that this result is also dependent on couplings in the scalar potential through $f_H$, but it is small enough to be negligible compared to the larger contributions for any choice of those couplings which is perturbative.

\subsection{$Z^{\prime}$ Boson}

Now let us consider the contribution of the new neutral gauge boson, which we denote as $Z'$. The only diagram which appears with this particle is in Figure \ref{vector1}. We note that the $Z'$ contribution is negative, pulling the overall result further away from the experimentally measured value. The result is given in Ref.~\cite{muonreport} as

\begin{eqnarray}
&&
\Delta a_{\mu} (Z^{\prime}(c)) = \frac{1}{8\pi^2}\frac{m_\mu^2}{ M_Z^{\prime 2} }\int_0^1 dx \frac{C^2_VP_V(x)+C^2_AP_A(x) }{(1-x)(1-\lambda^2 x) +\epsilon^2 \lambda^2 x},
\label{Ucontri2}
\end{eqnarray}where $C_V=-g\sqrt{3h_W}/4c_W$ and $C_A=g\sqrt{3h_W}/4c_W$ are the couplings between the muon and the $Z^{\prime}$ according to Eq.(\ref{4.9}), with $\epsilon \equiv 1$,\, $\lambda = m_{\mu}/M_{Z^{\prime}}$ and

\begin{eqnarray}
P_V(x) & = & 2 x (1-x)\cdot x \nonumber\\
P_A(x) & = & 2 x(1-x)\cdot (x-4)- 4\lambda^2 \cdot x^3.
\end{eqnarray}with $h_W = 1-4 s^2_W$ and $c_W=\cos\theta_W$.

These integrals simplify to give a contribution of 

\begin{equation}
\Delta a_{\mu}(Z^{\prime}(f)) = \frac{m_{\mu}^2}{4 \pi^2 M_Z^{\prime 2}}\frac{1}{3}\left(C^2_V - 5C^2_A\right).
\label{NPGZ}
\end{equation}

This is the contribution of the $Z^{\prime}$ to the muon anomaly magnetic moment.

\subsection{Doubly Charged Vector}

The doubly-charged boson, similarly to the doubly-charged scalar, gives rise to two diagrams that contribute to the $(g-2)_\mu$. The first one, shown in Fig. \ref{vector1}, is similar to the singly-charged gauge boson, with two differences: a multiplying factor of 4 due to the symmetry factors arising from identical fields in the interaction term, and an additional factor of 2 arising from the larger charge of the boson \cite{MuonC}.
\begin{eqnarray}
\Delta a_{\mu} (U^{\pm \pm})(a) &=&8\times \frac{f^2_U}{8\pi^2}\left( \frac{m_\mu}{M_{U^{\pm \pm}}}\right)^2\int_0^1 dx \frac{P_V(x) + P_A(x) }{\epsilon^2 \lambda^2(1-x)(1-\epsilon^{-2}x) + x }
\label{Vcontri}
\end{eqnarray}where $f_U=g/\sqrt{2}$ is the coupling strength between the muon and the new boson given in Eq.(\ref{4.6}), with $\lambda = m_{\mu}/M_{U^{\pm \pm}}$, and

\begin{eqnarray}
P_V(x) & = & 2 x^2(x-1) \nonumber\\
P_A(x) & = & 2 x^2(x+3)+4 \lambda^2 \cdot x (1-x)(x-1).
\end{eqnarray}

These integrals simplify to give

\begin{eqnarray}
\Delta a_{\mu}(U^{\pm \pm})(a) = \frac{2f_U^2 m_{\mu}^2}{\pi^2 M_{U^{\pm \pm}}^2}\left( \frac{10}{6}\right).
\label{UppEq1}
\end{eqnarray}

The second diagram, shown in Figure \ref{vector2}, is similar to the $Z^{\prime}$ one, but we once again have a factor of 4 due to the identical fields, and we also have a relative negative sign due to the opposite charge of the muon running in the loop. Therefore we find,

\begin{eqnarray}
&&
\Delta a_{\mu} (U^{\pm \pm})(b) =(-4)\times \frac{1}{8\pi^2}\left( \frac{m_\mu}{M_{U^{\pm \pm}}}\right)^2\int_0^1 dx \frac{P_V(x)+P_A(x) }{(1-x)(1-\lambda^2 x) + \epsilon^2 \lambda^2 x}
\label{Ucontri2}
\end{eqnarray}with $\epsilon \equiv 1$, $\lambda = m_{\mu}/M_{U^{\pm \pm}}$, and

\begin{eqnarray}
P_V(x) & = & 2 x (1-x)\cdot x \nonumber\\
P_A(x) & = & 2 x(1-x)\cdot (x-4)- 4\lambda^2 \cdot x^3.
\end{eqnarray}

This second contribution simplifies to

\begin{equation}
\Delta a_{\mu} (U^{\pm \pm})(b) = \frac{4m_{\mu}^2 f^2_U}{3 \pi^2 M_{U^{\pm \pm}}^2}.
\label{NPGUb}
\end{equation}

The total doubly charged boson contribution is given by

\begin{eqnarray}
\Delta a_{\mu} (U^{\pm \pm})(Total)&=& \Delta a_{\mu} (U^{\pm \pm})(a)+\Delta a_{\mu} (U^{\pm \pm})(b)\\
&=& \left( \frac{7}{2}\right) 
\left( \frac{f_U m_{\mu}}{\pi M_{U^{\pm \pm}}} \right)^2.
\label{doublyvector}
\end{eqnarray}

\subsection{Neutral Scalars}

For the new neutral scalar the only diagram of relevance is shown in Figure \ref{scalar1}, which gives an irrelevant contribution to the $(g-2)_\mu$ in agreement with \cite{PDG,g2muon331_3,g2muon331_4,muonreport,muonBSM}. Below we present the analytical expressions for the Higgs and heavy Higgs contributions,

\begin{eqnarray}
&&
\Delta a_{\mu} (h) = \frac{1}{8\pi^2}\frac{f^2_h m_\mu^2}{ M_h^2 } \int_0^1 dx \frac{P_S(x) }{(1-x)(1-\lambda^2 x) +\epsilon^2 \lambda^2 x}
\end{eqnarray}where 

\begin{eqnarray}
P_S(x) & = &  x^2 (1+\epsilon -x)
\end{eqnarray}which gives us,

\begin{eqnarray}
&&
\Delta a_{\mu} (h) = \frac{1}{8\pi^2}\frac{f^2_h m_\mu^2}{ M_h^2 } \left[ 2\ln \left(\frac{M_h}{m_{\mu}} \right) -\frac{14}{12}\right]
\label{hcontri}
\end{eqnarray}where,

\begin{equation}
f_h= \frac{m_l}{v_{\rho}}\left(\cos(\beta) - \dfrac{v_\rho}{v_{\chi}}\sin_(\beta) \right).
\end{equation}

Similarly for the heavy Higgs we find

\begin{eqnarray}
&&
\Delta a_{\mu} (h) = \frac{1}{8\pi^2}\frac{f^2_h m_\mu^2}{ M_{h^2_2} } \left[ 2\ln \left(\frac{M_{h_2}}{m_{\mu}} \right) -\frac{14}{12}\right]
\label{hcontri2}
\end{eqnarray}with

\begin{equation}
f_{h_2} = \frac{m_l}{v_{\rho}}\left(\sin(\beta) + \dfrac{v_\rho}{v_{\chi}}\cos(\beta) \right).
\end{equation}

\begin{figure}[!h]
\centering
\includegraphics[scale=1]{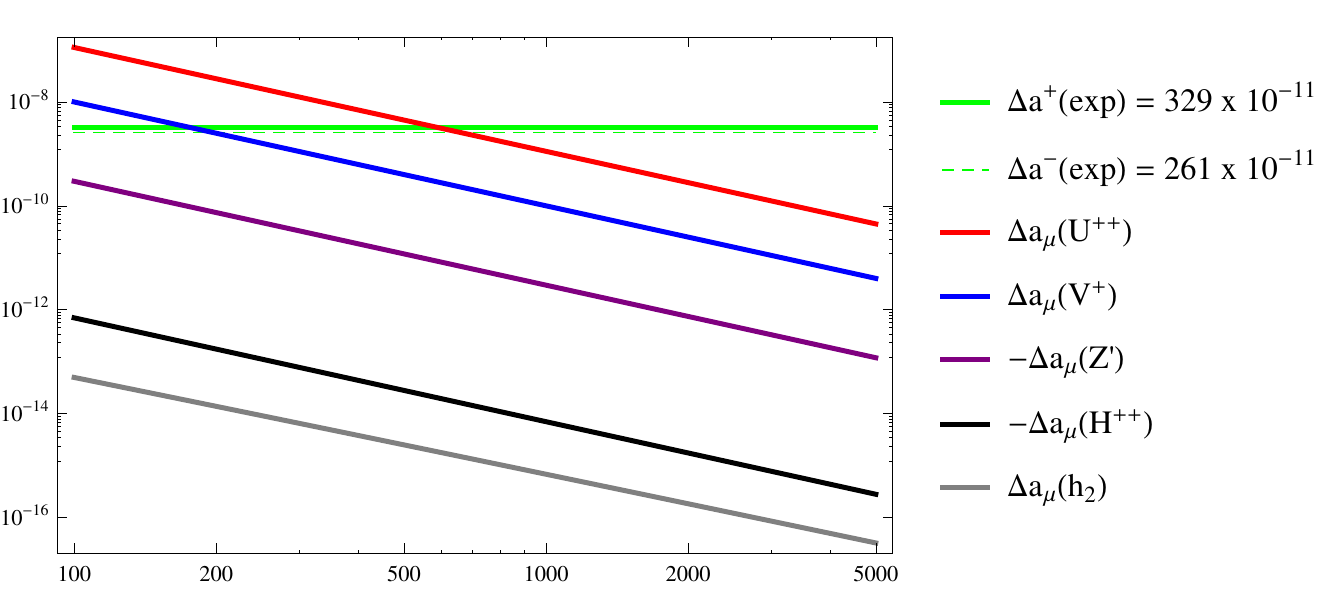}
\caption{Contributions from each new particle in the RM331 model to the anomalous magnetic moment of the muon, plotted against that particle's mass. The projected range from the Fermilab experiment is shown with the upper (lower) 1$\sigma$ value in solid (dashed) green lines. The current range is larger by about a factor of 2. Note that the contribution of the $Z'$ boson and the $H^{\pm\pm}$ is negative. The strong hierarchy between the contributions means that the total correction lies just above the uppermost curve ($U^{\pm\pm}$).}
\label{resultvsmass}
\end{figure}

\begin{figure}[!h]
\centering
\includegraphics[scale=1]{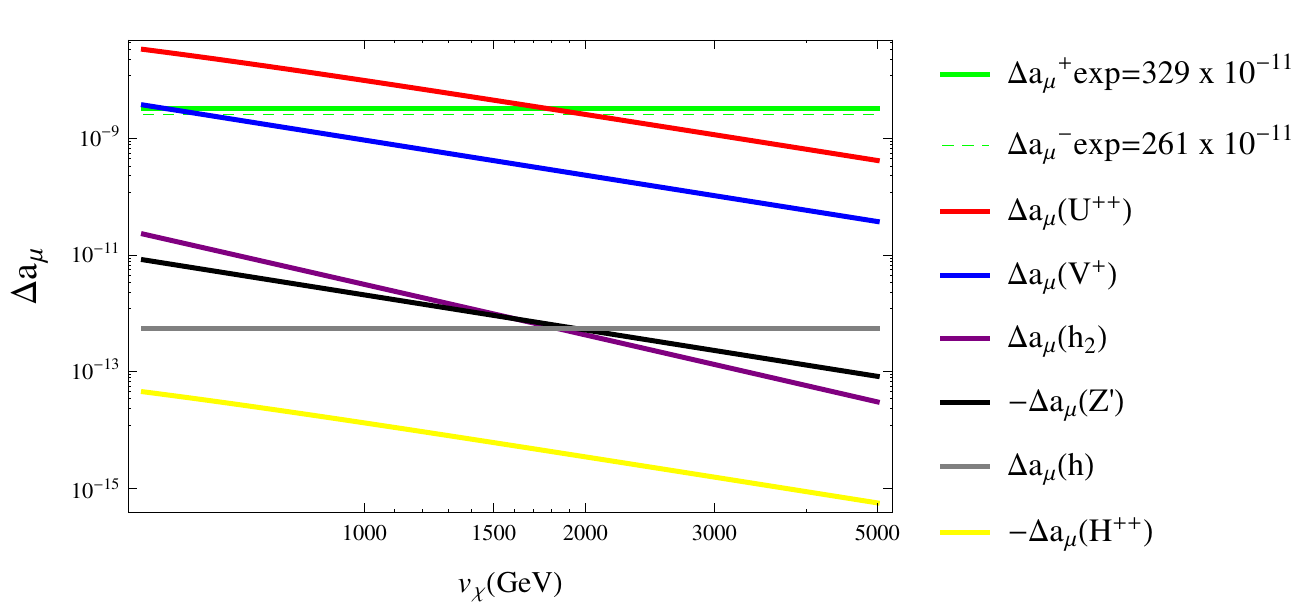}
\caption{The contributions from each of the new bosons in the RM331 model are labelled in the figure, plotted against the symmetry-breaking scale of the model, which sets the masses of the new particles. The projected experimental range is again shown in solid/dashed green lines. We have assumed a value of $\lambda_{4}=1$ in calculating the $H^{\pm\pm}$ contribution. The contribution is sensitive to this choice quadratically, indicating that the $H^{\pm\pm}$ contribution is small for any perturbative choice of parameters. We can conclude that if the discrepancy in $a_\mu$ is to be explained by the new physics of the RM331 model, we require a symmetry breaking scale of approximately $2$~TeV, and therefore with $M_{V^{\pm}} \simeq 650$~GeV, $M_Z^{\prime} \simeq 2.4$~TeV and $M_{U^{\pm \pm}} \simeq 660$~GeV}.
\label{result}
\end{figure}

All of the above contributions given in Eqs.(\ref{NPGa}), (\ref{doublyScalar}), (\ref{NPGZ}), (\ref{doublyvector}), (\ref{hcontri}), (\ref{hcontri2}), are displayed in Figure \ref{resultvsmass} as a function of the, assumed equal, mass of the new particle, and in Figure \ref{result} as a function of the $SU(3)_L\otimes U(1)_N$ symmetry breaking scale, which sets the mass of these new particles through spontaneous symmetry breaking. The second plot is more physical, as the masses are all correlated but not identical. In fact, we can see that some of the contributions evolve differently with symmetry breaking scale than others. We find that, if the discrepancy in $a_\mu$ is to be explained by the new physics of the RM331 model, we require a symmetry breaking scale of approximately $1.7-2$~TeV and hence favoring $ 555\ \mbox{GeV} \lesssim M_{V^{\pm}} \lesssim 652$~GeV, $2.35\ \mbox{TeV} \lesssim M_Z^{\prime} \lesssim 2.4$~TeV and $600\ \mbox{GeV} \lesssim M_{U^{\pm \pm}} \lesssim 657$~GeV.

\begin{figure}[!h]
\centering
\includegraphics[scale=0.8]{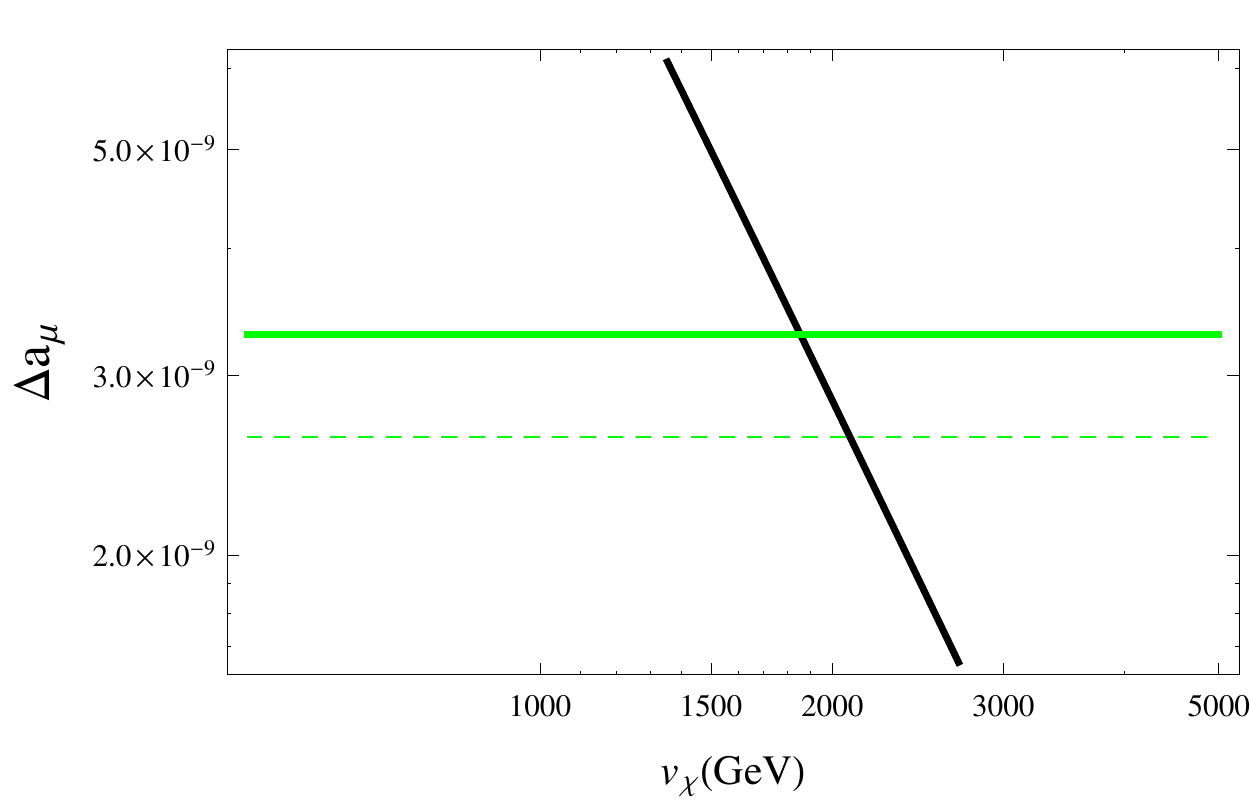}
\caption{Total contribution of the reduced minimal 331 model to the $(g-2)_{\mu}$ along with the expected 1 sigma range from the Fermilab experiment. Notice that a scale of symmetry breaking at $\sim 2$~TeV is favored. The current $2\sigma$ upper bound for the breaking scale is $1.5$~TeV and the Fermilab experiment should improve this to 1.7~TeV.}
\label{total}
\end{figure}
 
For completeness, we show the total contribution of the reduced minimal 3-3-1 model to the anomalous magnetic moment of the muon as a function of the scale of symmetry breaking in Fig.~\ref{total}. Our results are somewhat consistent with previous works on this topic \cite{g2muon331_1,g2muon331_2,g2muon331_3,Yue:2013qba}. We did not restrict ourselves to one particular sector as done in previous works and we have presented analytical expressions for all leading order contributions.

\section{Conclusions}
\label{sec4}

We have calculated all the leading-order contributions to the muon anomalous magnetic moment in the Reduced Minimal 3-3-1 model. We find that this model can reproduce the current experimental results, but only for a very narrow window of symmetry breaking scales ($1.7-2$~TeV) if the results of the new run of the $a_\mu$ experiment at Fermilab give the expected precision at the current experimental central value. This leads to very specific predictions for LHC physics, predicting light gauge bosons with $ 555\ \mbox{GeV} \lesssim M_{V^{\pm}} \lesssim 652$~GeV, $2.35\ \mbox{TeV} \lesssim M_Z^{\prime} \lesssim 2.4$~TeV and $600\ \mbox{GeV} \lesssim M_{U^{\pm \pm}} \lesssim 657$~GeV. Some bounds on these gauge bosons have already been derived using LHC data \cite{LHCbosons1,LHCbosons2,LHCbosons3}. The authors focused on a different 3-3-1 model, however, and the translation between these models is nontrivial. We expect LHC bounds to be similar and therefore to offer a complementary bound to this one based on $a_\mu$.

If we instead suppose that new experimental or theoretical results resolve the anomaly in $a_\mu$, we can then place a lower bound on the scale of $SU(3)_L\otimes U(1)_N$ breaking by requiring that the contribution to the anomalous magnetic moment not be above the error in the measured value. Applying this criterion, the current lower bound for the breaking scale is 3.75~TeV by taking  $\sigma\left(a_{\mu}\right)= 81 \times 10^{-11}$.  This value should increase to 5.8~TeV with the proposed Fermilab experiment and improvements in the calculation to the SM contribution which predict $\sigma\left(a_{\mu}\right)= 34 \times 10^{-11}$. Given the constraints imposed by the Landau pole concerns discussed earlier, this second bound would rule out the theory in its entirety. These bounds could only be evaded by a finely-tuned conspiracy of cancellations between the contributions of the RM331 model and some other physics which also affects $a_\mu$.

\section{Acknowledgements}
FQ is partly supported by the Brazilian National
Counsel for Technological and Scientiﬁc Development (CNPq). WS is partly supported by the U.S.\ Department of Energy under contract DE-FG02-04ER41268. PP thanks the hospitality of UFMA during the stages of this work.

\end{document}